\documentclass[preprint,journal]{vgtc}            

\onlineid{}

\preprinttext{\copyright IEEE. This is the author's version of the article. To appear in IEEE Transactions on Visualization and Computer Graphics.}

\vgtccategory{Research}

\title{Non-urgent Messages Do Not Jump into My Headset Suddenly! \\ Adaptive Notification Design in Mixed Reality}

\author{%
  \authororcid{Jingyao Zheng}{0000-0001-8920-308X},
  \authororcid{Xian Wang}{0000-0003-1023-636X},
  \authororcid{Sven Mayer}{0000-0001-5462-8782} and 
  \authororcid{Lik-Hang Lee}{0000-0003-1361-1612}
}

\authorfooter{
  \item
  	Jingyao Zheng, Xian Wang, and Lik-Hang Lee are with Hong Kong Polytechnic University.
  	E-mails: jingyao.zheng@connect.polyu.hk, xiann.wang@connect.polyu.hk, and lik-hang.lee@polyu.edu.hk
  \item
  	Sven Mayer is with TU Dortmund University.
  	E-mail: info@sven-mayer.com.
  \item 
    Lik-Hang Lee is the corresponding author.
}

\abstract{
    Mixed reality (MR) notification systems currently display all messages in fixed central locations regardless of urgency, leading to unnecessary interruptions and cognitive overload. Drawing from previous MR/Virtual Reality (VR) notification design work and calm technology principles, we developed an adaptive notification system that adjusts spatial placement based on urgency levels: non-urgent notifications appear as peripheral icons accessible via head movement, moderately urgent messages anchor to the user's hand, and very urgent notifications transition progressively from peripheral to central view. Through a within-subjects study (N=18), we evaluated our adaptive system against the default centralised approach. Results demonstrate that the adaptive system significantly reduces mental workload ($p=0.041$), temporal workload ($p=0.008$), and frustration ($p=0.004$) while maintaining comparable notification awareness. Logistic regression analysis reveals that users prefer the adaptive system even with classification errors, provided the combined misclassification rate (disruptiveness + omission errors) remains below a determinable threshold. Our findings establish the first empirical evidence that urgency-based spatial notification distribution effectively addresses core MR usability challenges, offering practical design guidelines for immersive notification systems that balance user attention management with information accessibility.
}


\teaser{
  \centering
  \includegraphics[width=\linewidth]{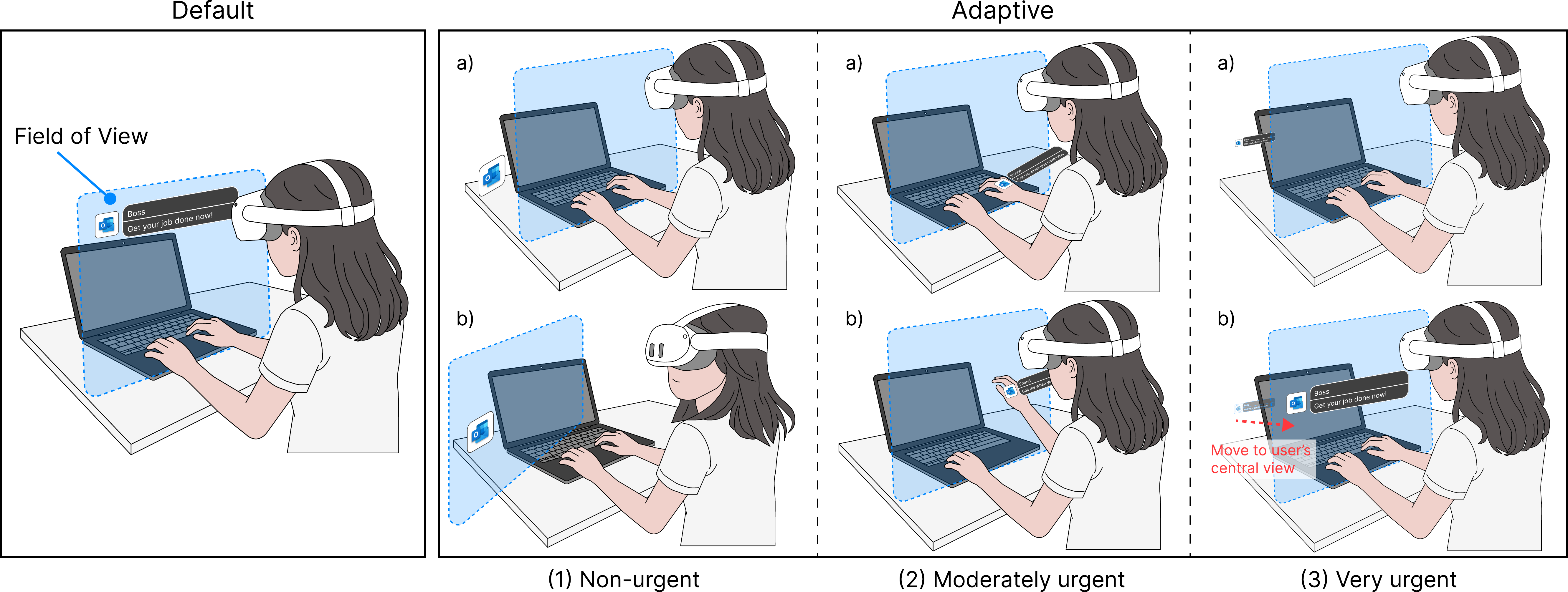}
  \caption{%
  	Overview of our adaptive notification user interface for Mixed Reality. Unlike the default system (Left), which anchors all notifications at the top of the field of view (blue area), our design adapts placement by urgency: (1) non-urgent notifications appear as small icons outside the central view with the easy access via slight head movement; (2) moderately urgent notifications follow the user’s hand for accessible yet non-disruptive awareness; and (3) very urgent notifications emerge at the visual periphery and smoothly transition into the user’s central view to ensure timely attention.%
  }
  \label{fig:teaser}
}

\graphicspath{{figs/}{figures/}{pictures/}{images/}{./}} 

\usepackage{tabu}                      
\usepackage{booktabs}                  
\usepackage{lipsum}                    
\usepackage{mwe}                       
\usepackage{amssymb}
\usepackage{mathptmx}                  
\usepackage{amsmath}
\usepackage[yyyymmdd]{datetime}

\usepackage{xcolor}
\usepackage{colortbl}
\usepackage{tabularx}
\usepackage{dcolumn} 
\newcolumntype{d}[1]{D{.}{.}{#1}}
\usepackage{multirow} 

\usepackage{todonotes}
\let\xtodo\todo
\renewcommand{\todo}[1]{\xtodo[inline,color=green!50]{#1}}

\definecolor{cellblue}{rgb}{0.67, 0.84, 0.90} 
\definecolor{cellred}{rgb}{1, 0.5, 0.5}
\definecolor{not urgent}{HTML}{A5B592} 
\definecolor{urgent}{HTML}{C82020}
\definecolor{moreately urgent}{HTML}{F6BB75}

\begin{document}



\firstsection{Introduction}

\maketitle


Mixed reality (MR) becomes more common in daily use, and users struggle to manage constant notifications without disrupting their main tasks or breaking immersion. In this paper, we use the term MR synonymously with Augmented Reality (AR) to refer to the overlay of virtual objects onto the physical world~\cite{speicher2019mixed}. While smartphones display notifications on fixed 2D screens, MR enables spatial notification placement throughout 3D space, creating both design challenges and opportunities that require strategic positioning for optimal user experience. Current MR notification systems in commercial headsets, such as Apple Vision Pro and Quest Pro, show all incoming messages in a fixed central location, regardless of their importance, potentially leading to unnecessary interruptions and overwhelming users with information~\cite{10.1145/3467963}. Similar challenges exist in Virtual Reality (VR), where poorly designed notifications break immersive experiences and create unwanted disruptions~\cite{hsieh2020bridging, ghosh2018notifivr}. However, completely isolating users from notifications would induce anxiety and disconnection from their digital lives~\cite{pielot2017beyond}. This creates a fundamental design tension: users need to stay connected to urgent information while maintaining focus on their primary MR tasks. Therefore, there is a critical need for adaptive, user-centred notification systems that can reduce notification disruptiveness while maintaining notification awareness.


Extensive studies on VR/MR notifications focused on multi-modal approaches~\cite{ghosh2018notifivr, marques2022which, plabst2025order} and User Interface (UI) strategies~\cite{imamov2020display, rzayev2020effects, hsieh2020bridging}. Studies on static notification placement have provided valuable insights. For example, Rzayev et al. explored four placement types and found that Head-Up Display was perceived as most urgent but least preferred, while On-Body and Floating placements achieved the highest preference and understandability scores~\cite{rzayev2019notification}. Recently researchers have begun exploring adaptive VR/MR notifications. Early work by Lindlbauer et al. developed a system with five detail levels for notification UIs based on cognitive load~\cite{lindlbauer2019context}, while recent studies have addressed gaze-based adaptation~\cite{kawakubo2025dynamic} and environmental context~\cite{li2024situationadapt}. However, existing adaptive notification research has not effectively integrated established spatial design principles from static placement studies into adaptive systems that dynamically adjust notification presentation. This gap necessitates developing and empirically validating adaptive notification systems that leverage proven spatial design knowledge for urgency-aware placement.


Our work addresses this gap by developing an adaptive notification system that adjusts presentation based on urgency levels. We focus on urgency rather than importance as our design criterion because urgency correlates strongly with users' acceptance of interruptions~\cite{vastenburg2009considerate}, whereas importance reflects long-term value without necessitating immediate response~\cite{Sahami2014large}. We define urgency as the willingness to respond to a notification at a given moment, capturing the behavioural intention aspect of receptivity that \textit{anticipates a user's subjective overall reaction to an interruption}~\cite{fischer2010Receptivity}. Based on Vastenburg et al's framework~\cite{vastenburg2009considerate}, we adopted a three-tier urgency system: Non-Urgent, Moderately Urgent, and Very Urgent. Each tier employs distinct spatial positioning strategies (see~\autoref{fig:teaser} and detailed in~\autoref{subsec: Adaptive Design}): non-urgent notifications are positioned outside users' periphery as application icons and accessible via slight head movement, moderately urgent notifications are hand-anchored to maintain awareness while minimising distraction, and very urgent notifications transition gradually from the peripheral left rear position to central view until acknowledged.


To evaluate the adaptive notification system's effectiveness and usability, we conducted a within-subjects study comparing it with the default system. Unlike prior work that assessed notifications within isolated task contexts~\cite{rzayev2019notification, hsieh2020bridging}, our study evaluated users' overall system preferences after experiencing continuous notification streams across different activities and environments. This approach better reflects real-world usage, where users evaluate notification systems based on their overall experience managing constant information flow rather than performance in any single context. We first conducted a pilot study (N=6) to refine system parameters, including notification positions and interactions, and ensure participant comfort with the notification presentations. This research investigates two primary questions: \textbf{RQ1:} Can our adaptive notification system, integrated with the previous notification design (e.g.,~\cite{ens2014cockpit, ghosh2018notifivr}) and attention management findings(e.g., ~\cite{matthews2007defining, stothart2015attentional}), effectively reduce users' cognitive load while maintaining notification awareness, even with classification errors? \textbf{RQ2:} How do user preferences change (adaptive notification system vs. default one) when classification errors occur?

Our study with 18 participants demonstrates that the adaptive system significantly reduces mental workload, temporal workload, and frustration levels compared to the default system (which displays all notifications at the top of users' central view) while maintaining comparable notification awareness. We establish the first empirical evidence that users prefer adaptive notification systems even when classification errors occur. Through multivariate logistic regression analysis, we found that user preference is significantly influenced by the combined classification accuracy, specifically the balance between \textit{omission} errors (urgent notifications misclassified as non-urgent) and \textit{disruptiveness} errors (non-urgent notifications incorrectly elevated to higher urgency levels). These findings provide crucial insights for AI-driven notification management systems: adaptive approaches remain beneficial even with imperfect urgency classification, provided the combined error rate stays below our regression-determined threshold. This relationship suggests that designers should focus on intelligently balancing these error types rather than pursuing perfect classification accuracy, offering practical guidance for developing robust notification systems in immersive environments.


\section{Related Work}

\subsection{Notifications in Immersive Environments}

Previous work showed that naively overlaying pop-up messages can break immersion in immersive environments~\cite{ghosh2018notifivr, hsieh2020bridging}. Accordingly, numerous studies in VR and MR have explored how and where notifications appear, which affects user experience and their focus. For example, previous research has examined the effectiveness of notifications delivered through various multimodal formats~\cite{marques2022which, plabst2025order, cho2025evaluating}. Ghosh et al.\cite{ghosh2018notifivr} investigated multi-modal approaches to VR notifications, examining visual, auditory, and haptic modalities and indicated that haptic notifications demonstrated the lowest effectiveness compared to visual and auditory alternatives, a conclusion corroborated by subsequent research conducted by George et al.\cite{george2020invisible}. Furthermore, George et al. examined how different notification modalities, such as text, ambient, and spotlight, affect users' sense of presence in virtual reality environments through interruption~\cite{george2018intelligent}. Their findings revealed that virtual text notifications had the most detrimental impact on presence compared to ambient and spotlight modalities.

From a visual perspective, Rzayev et al. systematically compared notification placements in VR, including a heads-up display, on-body cues, world-anchored panels, and in-situ object-anchored notifications~\cite{rzayev2019notification}. They demonstrated that placement significantly shaped perceptions of noticeability, intrusiveness, and urgency, with messages in a persistent heads-up display perceived as the most urgent. Similarly, Hsieh et al. tested three types of notification placement, including head-mounted, controller-anchored and movable panel displays, finding that suitability depends on activity pace and interference with ongoing interactions~\cite{hsieh2020bridging}. For example, notifications that appear in the same area where users are actively interacting feel poorly timed and disruptive. Additionally, Rzayev et al. specifically explored the position and alignment of notifications in social contexts and found that centre-positioned notifications with receiver-locked alignment created higher perceived urgency~\cite{rzayev2020effects}. However, existing research has primarily compared static notification approaches rather than developing adaptive systems that intelligently adjust to varying urgency levels in real-time.  Our work addresses this gap by integrating established design principles into a unified adaptive system that dynamically responds to notification urgency.

\subsection{Dynamic and Adaptive Notification Interfaces}

Early dynamic systems in MR/VR have focused on three key aspects of UI adaptation: placement, timing, and information density. Lindlbauer et al. developed an optimisation-based controller that monitors factors such as cognitive load and current activity, then continuously reconfigures MR interfaces by reducing UI elements or detail levels as cognitive demand increases~\cite{lindlbauer2019context}. Building on spatial adaptation, Li et al. introduced \textit{SituationAdapt}, which optimises UI placement based on environmental and social contextual cues~\cite{li2024situationadapt}. Furthermore, Li et al. employed long short-term memory (LSTM) networks to predict user noticeability in VR environments and dynamically adapt UIs accordingly~\cite{li2024predicting}. Recent research has specifically targeted notification systems, with Kawakubo et al. developing methods to relocate notifications relative to users' gaze clusters, balancing salience with interference reduction~\cite{kawakubo2025dynamic}. Complementing these approaches, Lu et al. introduced the Glanceable AR paradigm, which enables efficient information acquisition through peripheral viewing on AR head-worn displays while reducing distracting or intrusive~\cite{lu2021Glanceable}. Furthermore, Ilo et al. introduced Goldilocks Zoning, a gaze-aware VR notification placement technique, that dynamically positions notifications in a ``just-right'' region of the user’s recent visual attention~\cite{GZ2024Ilo}. While these adaptive approaches have addressed cognitive load, environmental context, and gaze patterns, limited work has explored how to systematically combine urgency levels with spatial placement optimisation for notification systems in MR, nor have they been empirically tested through comprehensive evaluation across multiple task scenarios to assess overall system effectiveness.

\subsection{Calm Technology and Attention Management}

Effective notification system design must consider fundamental principles of human attention and cognitive load management. Calm Technology principles emphasise that technology should require minimal attention while engaging both central and peripheral awareness as needed~\cite{weiser1996designing}. Empirical research validates these design approaches: Stothart et al. demonstrated that smartphone notifications reduce task performance even without user interaction, highlighting the cognitive cost of notification presence~\cite{stothart2015attentional}. Perceptual load theory further predicts that under high attentional demand, minor visual distractors compete for scarce cognitive resources, emphasising the need for context-sensitive systems~\cite{lavie1995perceptual}. 
Ambient displays provide a practical implementation framework for these principles. Matthews et al. showed that peripheral displays maintain awareness while minimising interference by presenting low-priority information in unobtrusive, glanceable forms~\cite{matthews2007defining}. Similarly, Ishii et al. demonstrated that subtle peripheral cues, particularly motion-based ones, effectively maintain awareness without disrupting primary activities~\cite{ishii1998ambientroom}. Attentional shift models confirm that redirecting attention incurs cognitive switching costs, supporting position-based notification strategies~\cite{posner1980orienting}.

Building on these foundations, our adaptive notification system applies calm technology principles by positioning notifications based on urgency levels, using peripheral placement for low-priority information and progressive disclosure for urgent content.

\section{Method}
Inspired by previous work~\cite{CalmTechnology_Amber, rzayev2019notification, ghosh2018notifivr, grubert2015multifi}, we developed an adaptive notification system that dynamically adjusts notification UIs based on urgency levels. To evaluate its usability and effectiveness, we conducted a within-subjects study examining two conditions (Adaptive vs. Default). The default notification system displays all notifications in the top area of the central view regardless of urgency~\cite{10.1145/3467963}, while our adaptive system varies presentation based on urgency classification. 

We designed the study to reflect continuous notification exposure across diverse daily activities because this mirrors how users actually encounter and evaluate notification systems in practice. Rather than optimising for specific task contexts, we assessed whether our adaptive system could effectively manage the continuous stream of notifications that characterises real-world MR usage. 

As such, our hypotheses are as follows: \textbf{[H1]:} Users will prefer the adaptive MR notification system over the default system even when classification errors occur. \textbf{[H2]:} User preference for adaptive MR notification interfaces will correlate positively with classification accuracy. \textbf{[H3]:} Despite classification errors, participants will experience lower cognitive workload with the adaptive position-based MR notification system compared to the default static system. \textbf{[H4]:} The adaptive MR notification system still maintains comparable or improved notification awareness despite the dynamic positioning approach.

Our hypotheses were grounded in the previous work's findings~\cite{gajos2008predictability, janaka2022paracentral, kawakubo2025dynamic, rzayev2019notification, hsieh2020bridging}. Gajos et al. demonstrated that adaptive UIs can enhance user satisfaction despite producing unpredictable behaviour (\textbf{[H1]}), and that satisfaction increases with prediction accuracy (\textbf{[H2]})~\cite{gajos2008predictability}. Furthermore, Janaka et al. showed adaptive positioning can reduce cognitive workload without degrading performance (\textbf{[H3]})~\cite{janaka2022paracentral}. Finally, based on extensive MR/VR notification placement studies~\cite{rzayev2019notification, hsieh2020bridging, kawakubo2025dynamic}, we expect users to maintain comparable notification awareness with dedicated placement design.

\subsection{Adaptive Notification Design}
\label{subsec: Adaptive Design}
Building on the urgency-based design framework established above, this section presents the specific notification designs and underlying design principles for each urgency tier.

\subsubsection{Non-Urgent Notification Design} 
Non-urgent notifications are defined as messages that contain information of low temporal priority and do not require user action during the current task or sub-task. They are positioned outside the user's immediate perceptual field (left) to minimise attentional costs and preserve immersion (see \autoref{fig:teaser} (1)). To maintain accessibility while addressing these concerns, our design clusters non-urgent notifications into source application icons accessible via slight head movement. By clicking the icons, users can check all upcoming notifications from the applications. This approach follows ambient display best practices that group low-priority information into unobtrusive, glanceable forms~\cite{matthews2007defining}, allowing users to check updates at self-chosen moments without cognitive overload.

\textbf{Design Principles:} This design decision aligns with Calm Technology principles, which emphasise that \textit{technology should require the smallest possible amount of attention}~\cite{CalmTechnology_Amber}, and is supported by substantial evidence that even small visual elements can disrupt task performance. Stothart et al. demonstrated that smartphone notifications reduced performance on attention-demanding tasks, even when users did not respond to or interact with the notifications~\cite{stothart2015attentional}, suggesting that simply knowing a notification is present can degrade performance. Similarly, perceptual load theory predicts that under high attentional demand, even minor distractors such as static icons compete for scarce cognitive resources~\cite{lavie1995perceptual}, while attentional shift models confirm that redirecting attention incurs switching costs~\cite{posner1980orienting}.

\subsubsection{Moderately Urgent Notifications} 

Moderately urgent notifications are defined as messages that require the user’s timely attention but do not demand immediate disruption of the ongoing primary task. These notifications are spatially anchored to the user's dominant hand (See \autoref{fig:teaser} (2)). Users can subsequently move their hand to relocate these notifications to their central viewing area when ready to engage with the content.

\textbf{Design Principles:} The interfaces should provide visibility while minimising interference with the user’s cognitive flow. Accordingly, moderately urgent notifications are placed on hand. Hands are considered natural information displays in spatial computing~\cite{ens2014cockpit, grubert2015multifi, plabst2022push}. Extensive empirical research on notification UI placement in immersive environments confirms that hand-anchored notifications are more natural and preferred by users~\cite{rzayev2019notification, ghosh2018notifivr, plabst2022push}. Theoretically, this design draws on principles of embodied cognition, which view the body as an integral component of the cognitive system~\cite{dourish2001action, wilson2002six}. In MR, the hands are constantly monitored for manipulation, pointing, and control, making them a natural attentional anchor. By situating notifications on the hand, the system reduces disruptive gaze shifts to unrelated areas of the visual field, while preserving awareness of upcoming events.

\subsubsection{Very Urgent Notifications}
Very urgent notifications are defined as messages that contain critical, time-sensitive information requiring immediate user awareness and potential action. They initiate at the peripheral left position and fly toward the central field of view (see \autoref{fig:teaser} (3)). During this transition, users can accelerate the notification's arrival through gaze targeting or manual gestures, immediately bringing it to the central view when ready for information. 

\textbf{Design Principles:} This design employs a two-stage attention mechanism that balances immediate awareness with controlled interruption. The approach utilises peripheral positioning, as Calm Technology emphasises that technology should \textit{engage both the centre and the periphery of our attention, and in fact moves back and forth between the two}~\cite{weiser1996designing}. Furthermore, research on ambient displays demonstrates that subtle peripheral cues (e.g., motion) can maintain user awareness while minimising interference with primary tasks~\cite{ishii1998ambientroom, wisneski1998ambient}. Additionally, Rzayev et al.'s findings show that floating displacement was one of the most preferred and comfortable placements~\cite{rzayev2019notification}. Finally, the motion-based interaction allows users to determine optimal timing for information access, thereby reducing annoyance and cognitive load~\cite{adamczyk2004if}. Building on these insights, the specific left-side placement leverages pseudoneglect theory, which identifies a left visual field advantage for detection and localisation~\cite{jewell2000pseudoneglect}.

\subsection{Notifications in MR Interfaces}

\subsubsection{Pilot Study}
To ensure notification interfaces were positioned comfortably and interactions were human-centered, we conducted pilot studies with six participants (4 females and 2 males), aged from 23 to 28 ($M = 24.2$, $SD = 1.94$), and each compensated $\sim12.80\space USD$. The primary aim was to verify and adjust system settings and study design elements, such as the notification initial position and motion speed of very urgent notifications. Participants performed the same procedures planned for the main study and were encouraged to report any unreasonable or user-unfriendly system settings or procedures. Simple adjustments were implemented immediately based on participant feedback, while complex issues were documented for later development. All feedback was systematically recorded and analysed to refine system settings and parameters. Participants then confirmed these adjustments, ensuring the study design was optimised for effectiveness and user satisfaction before the main evaluation.

During the pilot study, we observed occasional gaze functionality malfunctions due to device constraints. Our study employed the Quest 3, which lacks eye-tracking capabilities. Consequently, our system relied on HMD-mounted cameras to approximate gaze direction. This approach caused recognition failures when users actively read notifications, but the cameras were not directly facing them, preventing the system from detecting user engagement through gaze. This led to the inclusion of gesture controls as an alternative interaction method to ensure robust user interaction regardless of device limitations. 

\subsubsection{UI Details} 
\begin{figure}[!t]
    \centering
    \includegraphics[width=\linewidth]{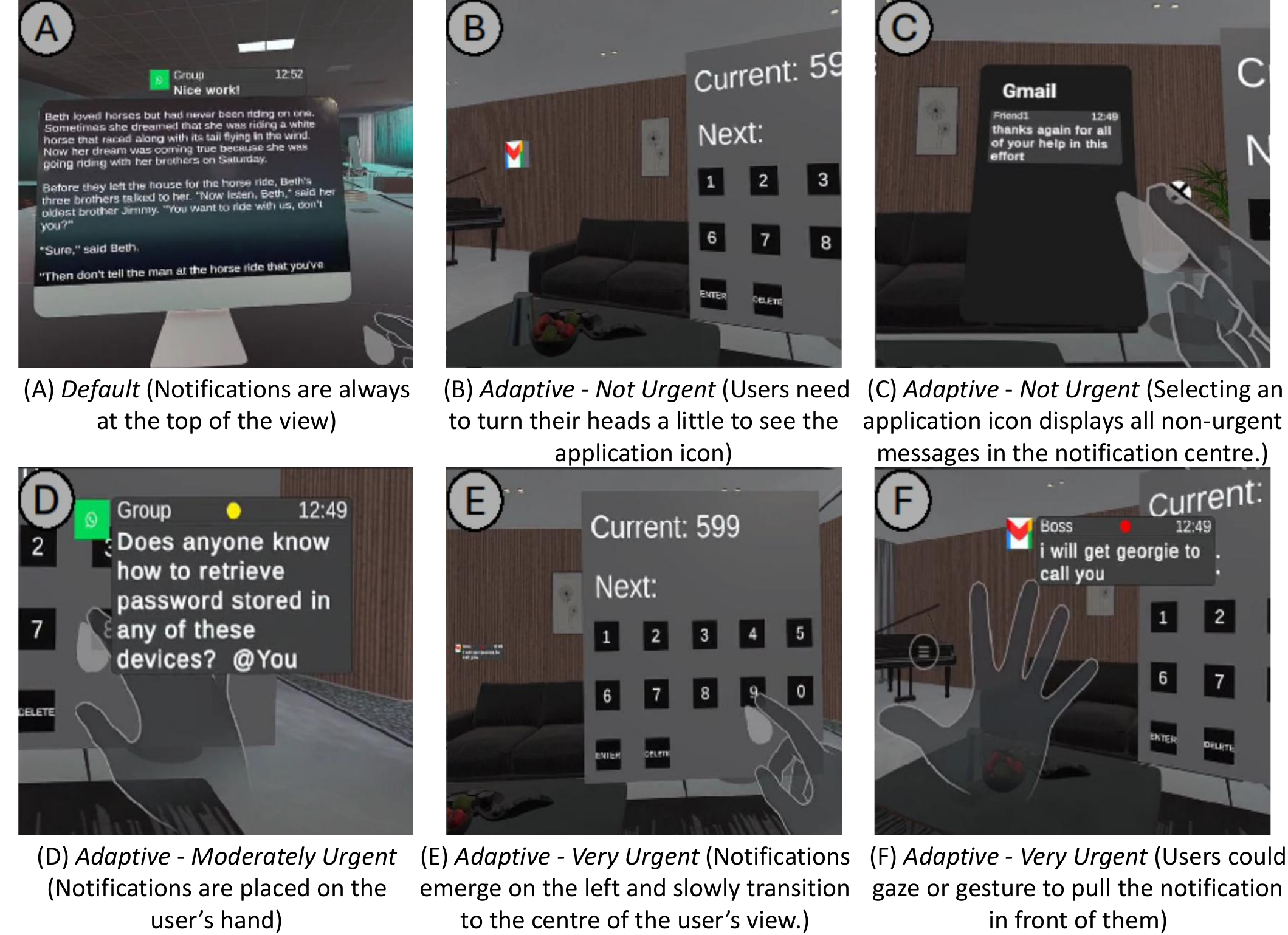}
    \vspace{-6mm}
     \caption{Notification Displacement}
    \label{fig:notification displacement}
    \vspace{-5mm}
\end{figure}

\autoref{fig:notification displacement} provides an overview of the two notification systems under \textit{default} and \textit{adaptive} conditions. Both systems employed expandable UI designs that automatically adjusted vertical dimensions based on message length. Text specifications included a 14-point font for the sender and timestamp, a 16-point font for the notification content, and $0.25 cm \times 0.25 cm$ icons. In line with prior work~\cite{rzayev2019notification, zheng2025persono}, both systems automatically dismissed notifications if users did not interact with them within 15 seconds, except for the notification icons (not urgent notifications). 

The \textit{default} notification system served as the baseline and aimed to simulate the notification system in current HMDs, such as Apple Vision Pro. It pushed all notifications, regardless of the sender and the content, in the top area of central vision~\cite{10.1145/3467963} (see \autoref{fig:notification displacement} (A)). Specifically, notifications are positioned 0.35 meters forward from the camera along its forward vector, with a 0.1-meter upward offset.

In contrast, the \textit{adaptive} notification system displayed notifications based on urgency levels, balancing user workload with notification importance. For all notification interactions, eye gaze detection was defined as the angle between the ray from the eye centre anchor and the notification $< 10^\circ$.

For \textit{non-urgent} messages, application icons were positioned outside the user's periphery but remained within a comfortable region (0.2 meters left via $60^\circ$ from the forward vector, with 0.02-meter horizontal spacing between multiple icons), requiring slight head movement to view them (\autoref{fig:notification displacement} (B)). The system automatically adjusts notification orientation toward the user every 10 seconds, while preserving relative positional relationships. However, when the user's gaze aligned directly with an icon, the icon remained stationary and reset the ten-second timer. This design ensured notification accessibility during movement, such as walking. The interactive icons provided visual feedback: clicking an icon triggered the full notification interface (\autoref{fig:notification displacement} (C)).

Compared to the \textit{not urgent} notifications, the \textit{moderately urgent} and \textit{very urgent} notifications were much more noticeable. If the message was \textit{moderately urgent}, the \textit{adaptive} notification system located the message on a user's dominant hand (0.1 meters above the hand; See \autoref{fig:notification displacement} (D)), which would follow the hand until it was dismissed.

Moreover, the \textit{very urgent} notifications initially appeared on the left side at coordinates(-1, 0, 1) relative to the camera (See \autoref{fig:notification displacement} (E)). These notifications undergo a smooth 2-second transition to the user's central field of view (0.3 meters forward from the camera position). Alternatively, if the user detected them earlier, they could immediately be positioned directly in front of the user (See \autoref{fig:notification displacement} (F)). To simulate user detection of the \textit{very urgent} message, we employed two interaction modalities: eye gaze and gesture. Gesture was defined as shown in \autoref{fig:notification displacement} (F). These two interaction methods were tested in the pilot study, where participants affirmed their intuitiveness and ease of use.

To dismiss notifications, users could either gaze at the UI or use a gesture as shown in \autoref{fig:Notification_Interaction_Example} (E - F). When the user reads the notification (gazes at it), the progress circle starts to fill (see \autoref{fig:Notification_Interaction_Example} (E)). If the user's gaze remains on the notification past a specific time threshold, the progress circle completes, resulting in the disappearance of the notification UI. We calculated this time threshold based on the content length using the formula: $\textit{Time Threshold} = \frac{\textit{Number of Words}} {\textit{Average Reading Speed}}$. Here, the \textit{Average Reading Speed} was defined as 2.83 words per second, consistent with the reading speed estimated for the \textit{learning} task, as detailed in Section~\ref{sec:ReadingSpeed}. In addition to the gaze, the user could also use the gesture (See \autoref{fig:Notification_Interaction_Example} (F)).

\subsection{Notification Dataset}
To create a realistic notification dataset, we randomly selected 55 messages from the Mobile Text Dataset~\cite{vertanen2021mining}. This dataset was chosen to simulate real-world messaging scenarios. Previous work indicated that both the sender and content are crucial in determining the perceived urgency of notifications~\cite{chang2019think ,li2023alert, mehrotra2015designingContentDriven}. To better emulate real-world experiences and create contexts for urgency evaluation, we assigned realistic sender placeholders (e.g., Boss, Friend 1, Group) and modified notifications for naturalness. For example, we added ``@You'' to a group chat message and incorporated temporal indicators such as ``Now''.

Additionally, we enriched our dataset by including everyday notifications from social and non-social applications, such as ``Your friend has posted a new story'' and ``Your parcel has been delivered'', bringing the total to 60 messages. Each message was categorised by application type (messaging, group messaging, email, social, and non-social) to reflect realistic notification distributions, following Pielot et al.'s findings~\cite{pielot2018dismissed}. This categorisation simulated an authentic daily notification experience, mirroring typical user interactions with various applications.

We acknowledge the inherent limitations of notification classification systems, recognising that urgency assessment cannot achieve perfect accuracy due to the dynamic nature of human behaviour and cognition~\cite{li2023alert}. Such misclassifications may make adaptive MR notification interfaces more disruptive than beneficial, undermining their intended advantages. To simulate realistic usage conditions while addressing this limitation, we pre-labelled the notification dataset with urgency levels prior to the experimental trials. We intentionally constructed the dataset to reflect ``general'' preferences to ensure ecological validity while enabling fair and consistent comparisons across participants. Two researchers independently assigned urgency ratings to each notification based on their preferences. Following McDonald et al.'s guidance that inter-rater reliability analysis is unnecessary for straightforward rating tasks~\cite{McDonald2019Reliability}, the researchers collaboratively resolved any minor discrepancies through discussion to finalise the dataset.

We used labelled data to maintain experimental control rather than relying on a real-time AI system, which introduces inaccuracies and uncertainty into the experiment. Thus, this enabled us to isolate spatial positioning effects and systematically examine how misclassification rates affect user preferences through regression analysis.



\subsection{Task Design: Mixed Reality Simulation}
\label{Study}
This experiment utilised a VR system to emulate the impact of notification systems on users within an MR setting. We leverage VR to create a controlled environment that provides high certainty in environmental factors, specifically designed for MR evaluation. This approach, known as \textit{Mixed Reality Simulation}, has been extensively used in previous MR experiments~\cite{lu2022exploring, lee2010role, li2019gaze}. One key advantage of the approach is that our tasks require a robust yet seamless experience between users and MR, while the existing commercial MR headsets, especially sensors and tracking systems, cannot deliver such experiences. Thus, we situated an MR environment in our VR scenarios as our experimental testbed. 

\subsubsection{Apparatus}
We used a Windows 11 laptop with an NVIDIA GeForce RTX 4080 Laptop GPU, the 13th Gen Intel(R) Core(TM) i9-13980HX, and 32GB RAM connected to the Quest 3. The software was developed and implemented in Unity (ver 2022.3.22f1).

\subsubsection{Virtual Environments}

Users were placed in three selected virtual environments for different tasks (detailed in Section~\ref{sec:ReadingSpeed}): an \textit{outdoor maze} (\textit{Open}), a \textit{living room} (\textit{Semi-Open}) and an \textit{office} (\textit{Closed}) (See \autoref{fig:Tasks} A - C), which covered common environments and were in line with the environment categories used by Rzayev et al.~\cite{rzayev2019notification}. Each virtual environment was specifically selected to provide an ecologically appropriate context for its corresponding task and remained fixed throughout the study (i.e., Gaming always occurred in the outdoor maze, Problem Solving in the living room, and Learning in the office). For the outdoor maze environment, where participants were tasked with collecting balls, we prepared a physical $8 \times 6.7$ meters room free of obstacles. To ensure participants' safety, the outdoor maze scene was built slightly smaller, measuring $7.5 \times 6$ meters, than the actual room size. 

\subsubsection{Primary Task}
\label{sec:ReadingSpeed}

\begin{figure}[!t]
    \centering
    \includegraphics[width=\linewidth]{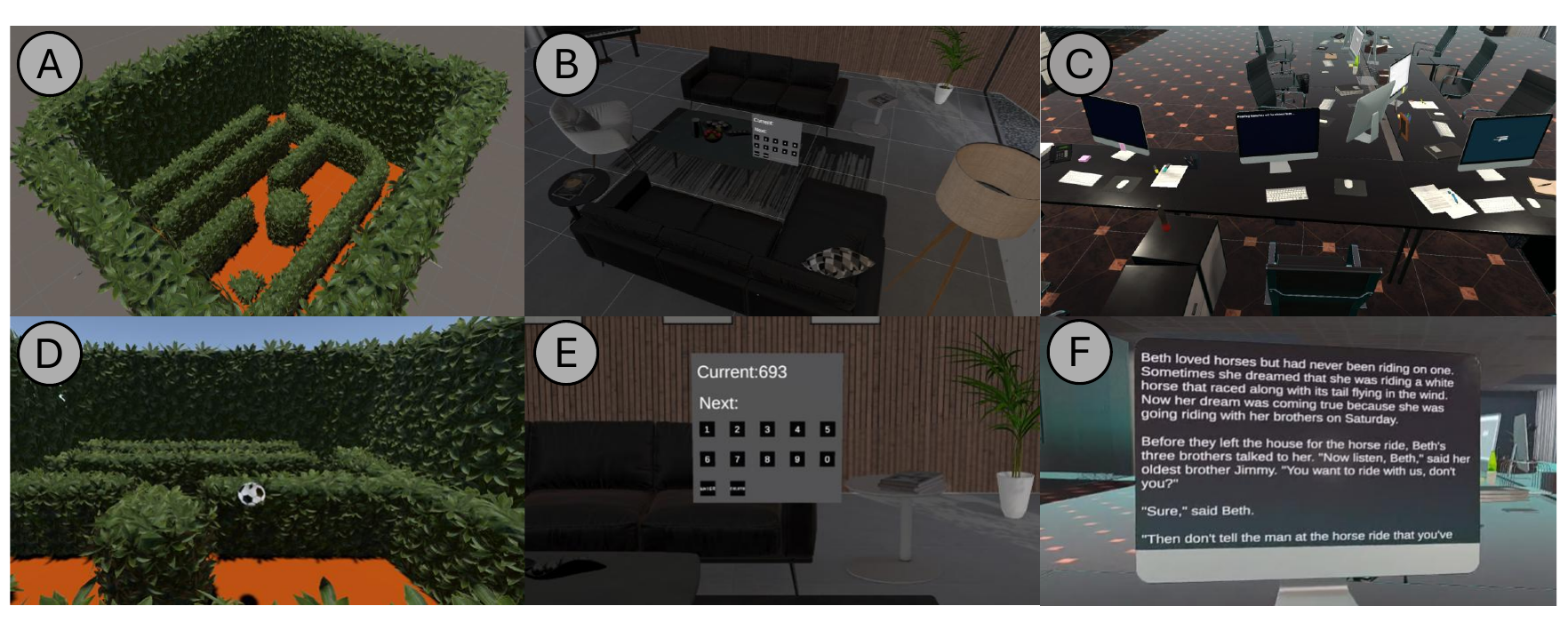}
    \vspace{-6mm}
    \caption{Three Environments in the Study:  (A) Maze; (B) Living Room; (C) Office; Three Primary Tasks in the Study: (D) \textit{Gaming} - Ball Collection; (E) \textit{Problem Solving} - Count 2; (F) \textit{Learning} - Reading Comprehension.}
    \label{fig:Tasks}
    \vspace{-7mm}
\end{figure}

For the primary tasks, we selected common VR use-cases and real-world activities: \textit{Gaming}, \textit{Learning}, and \textit{Problem Solving}~\cite{huang2010task}. Rather than selecting activities based purely on cognitive load theory, we prioritised ecological validity by choosing common daily activities that naturally vary in cognitive demands. This approach ensures our findings better reflect users' actual experience and evaluation of notification systems in realistic MR usage scenarios, enhancing the real-world applicability of our results. We set each primary task to last approximately 5 minutes for three key reasons. First, this duration ensured participants had sufficient time to engage deeply with each task. Second, many real-world tasks susceptible to notification interruptions, such as reading emails, writing brief reports, or completing design revisions, can be reasonably simulated within 5 minutes. Third, this duration balanced being sufficiently challenging to require focused attention while avoiding fatigue or concentration loss.

For the \textit{Gaming} task (See \autoref{fig:Tasks} (D)), we were inspired by previous work~\cite{rzayev2019notification, ghosh2018notifivr} and involved a ball collection task in the outdoor maze. Participants needed to walk around the maze and avoid obstacles to collect balls within five minutes, and they were instructed to collect as many balls as possible. Balls would randomly spawn in the environment at a fixed height, 1m above the floor, the same as the setting in Ghosh et al.'s work~\cite{ghosh2018notifivr}. For the \textit{Problem Solving} task (See \autoref{fig:Tasks} (E)) within the \textit{living room} environment, we selected the \textit{Count 2} task. Also, this is a common task in the human-computer interaction research~\cite{duchowski2018index, lindlbauer2019context}. Participants were tasked with counting two to the given number and inputting the correct result within 5 minutes. For the \textit{Learning} task (See \autoref{fig:Tasks} (F)), we chose a reading comprehension task using materials from EasyCBM~\cite{alonzo2006easycbm}\footnote{\href{https://www.easycbm.com/}{https://www.easycbm.com/}}, a resource widely used in research~\cite{chen2023easyCBMCHI, joshi2024easyCBMCHI}. Participants were provided a virtual keyboard to navigate pages and a ``Finished'' button on the last page to access the Q\&A section with answer buttons and a grabbable cheat sheet covering the reading material.

To maintain a five-minute study duration, we selected K2-grade materials of approximately 500 words based on three considerations: (1) college students read about 200 words per minute when preparing for multiple-choice tests~\cite{carver1992reading}; (2) second-language speakers read approximately 17\% slower than native speakers~\cite{cop2015reading, brysbaert2019reading}; and (3) decision-making in MR takes about 10\% longer than on desktop displays~\cite{rau2018speed}.
Given our non-English-speaking research context, we estimated participants would read at approximately 170 words per minute, completing the passage in about three minutes. Two minutes were reserved for notification interaction and question answering.

\begin{figure}[t!]
    \centering
    \includegraphics[width=\linewidth]{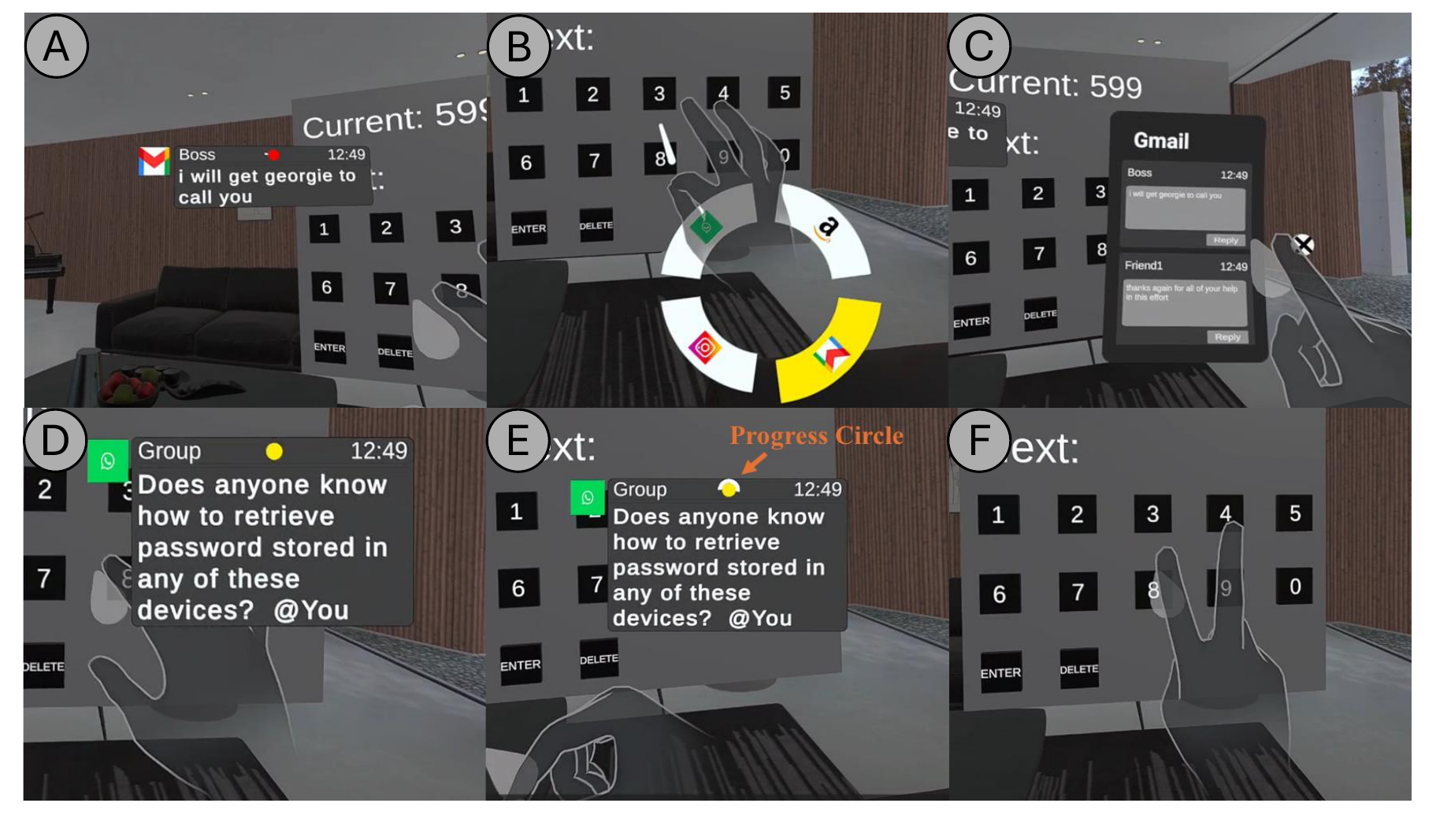}
    \vspace{-6mm}
    \caption{Examples of Notification Interaction: (A) Receiving and Reading a Very Urgent Message; (B) Poking for 1.5 Seconds to Open the Application Panel; (C) Interacting with Application UI; (D) Normal Notification UI; (E) \textit{Gaze} - Normal Notification UI with Loading Progress Circle; (F) \textit{Gesture} - Notification UI Disappear.}
    \label{fig:Notification_Interaction_Example}
    \vspace{-6mm}
\end{figure}

\subsubsection{Secondary Task}
During each task, ten notifications were randomly selected from our dataset to simulate realistic usage. In line with prior work~\cite{rzayev2019notification, zheng2025persono}, we delivered them at random intervals between 20-30 seconds, with no content repetition, to prevent anticipation effects. We tasked participants with viewing notifications and assessing the urgency levels of each message. If they deemed a message \textit{not urgent}, they could dismiss the notification without any further action. However, if they considered a message \textit{moderately urgent}, they were required to open the corresponding applications and interact with the message, i.e., clicking the ``Reply'' button. This interaction was to be done after completing a sub-task, such as entering a complete answer in the \textit{Count 2} task or finishing the current page of reading material, simulating real-world conditions. For messages perceived as \textit{very urgent}, participants needed to stop their ongoing work and interact with notifications immediately. 

We emphasised to users that they should not feel compelled to agree with the system's classifications, and they should always determine the notifications' urgency levels by themselves. For example, if a notification labelled \textit{very urgent} seemed \textit{not urgent} to them, they were justified in ignoring it. This guidance ensured the data reflected participants' true perceptions rather than compliance with system assessments.

\autoref{fig:Notification_Interaction_Example} provides an example of the interaction in which participants engaged with the notifications. First, they received a message classified as \textit{very urgent} by the adaptive notification system; see \autoref{fig:Notification_Interaction_Example} (A). After reading the notification, they agreed that this message is \textit{very urgent}, then they poked to trigger the application panel and opened the application by moving the hand to the relative position, i.e., Gmail; see \autoref{fig:Notification_Interaction_Example} (B). Once the application UI showed, they found the previous message and clicked the ``Reply'' button; see \autoref{fig:Notification_Interaction_Example} (C).


\subsection{Procedure} 
The entire user study lasted approximately two hours. After welcoming the participants, we outlined the study procedures and required them to sign an informed consent form and complete demographic questions. Before the main study, we clearly explained the definitions for \textit{non-urgent}, \textit{moderately urgent} and \textit{very urgent} notifications. To establish a baseline for which messages participants perceived as \textit{not urgent}, \textit{moderately urgent}, and \textit{very urgent}, we asked them to categorise the urgency levels of 90 messages. This initial classification was crucial for subsequent analyses of classification accuracy and message oversight. Sixty of these notifications were consistent with the dataset used during the study, but the content was paraphrased using an AI-based paraphrase tool, QuillBot\footnote{\href{https://quillbot.com/paraphrasing-tool}{https://quillbot.com/paraphrasing-tool}}, and subsequently verified by two of our researchers. The purpose of preparing 50\% more data and paraphrasing messages was to mitigate potential biases that could arise from participants' familiarity with the dataset. This approach ensured that each notification remained true to its intended urgency level but appeared linguistically unique to the participants.

After collecting the basic data, we presented an overview of how to interact with our notification systems and tasks via PowerPoint slides. Considering some participants might be very inexperienced or inexperienced with VR, we built an introductory scene in VR that allowed participants to practice interactions, such as poking to open the application panel (see \autoref{fig:Notification_Interaction_Example} (B)) and gesturing to dismiss the notifications. Once participants became familiar with the interactions, we reminded them to carefully read the sender and content, and independently determine urgency levels. Additionally, we thoroughly reviewed all requirements and details for both primary and secondary tasks to ensure participants fully understood their responsibilities during the study. Following the introductory segment, participants engaged with all \textsc{Task} $\times$ \textsc{Notification System} pairings in a sequence determined by Latin-square counterbalancing. Before concluding each session, participants verified whether any notifications remained unaddressed, particularly those that might have been missed in the application icon. Each task was designed to last approximately five minutes, with a three-minute rest period provided after each task to mitigate fatigue. After completing all the pairs, participants were asked to fill out the post-hoc questionnaires. 

\subsection{Participant}

Our university’s ethics board approved the study. In the main study, we recruited 18 participants (14 females and 4 males) by posting the poster on social media using our lab's official account. Their age ranged from 19 to 32 ($M = 23.4$ and $SD = 3.35$). All of them had either normal or corrected-to-normal vision and could view notification details clearly. Regarding their VR experience, five participants were \textit{inexperienced} with using VR, and six reported they were \textit{very inexperienced (i.e., I have only used VR once or twice before, if at all)}. Only one participant was \textit{experienced with using VR} and one was \textit{very experienced with using VR (i.e., I use VR several times a month)}. The rest of the five had \textit{some experience} with using VR. Concerning notification disturbance, two participants reported being \textit{very often} disturbed by mobile notifications, eight reported being \textit{often} disturbed, three felt \textit{occasionally} disturbed, and the final five were \textit{seldom} disturbed by notifications. It is noticeable that none of them were \textit{very seldom} disturbed by mobile notifications. We compensated participants with \$12.8 USD. 

\subsection{Measurement}
Our study primarily examined the alignment between users' perceived notification urgency and pre-labelled classifications. We marked notifications as \textit{misclassified} in three scenarios: when pre-labelled \textit{moderately urgent} messages weren't interacted with within 30 seconds (including a 15-second buffer), when \textit{very urgent} messages weren't addressed within 15 seconds, or when pre-defined \textit{non-urgent} notifications were interacted with within 30 seconds. 

We collected subjective data to assess user perceptions of both notification systems using standardised measures: raw NASA-TLX~\cite{hart2006nasa} for workload and SUS~\cite{brooke1996sus} for usability. Similar to the previous work~\cite{rzayev2019notification}, we have adapted the questionnaires about the Notification Mechanism originally from Weber et al.~\cite{weber2016design}. The questionnaire covers five dimensions with 5-point Likert scales as the following: (1) I'd feel comfortable using this notification mechanism when I am doing a task alone (\textit{Comfort}); (2) I'd feel comfortable using this notification mechanism when I am doing a task with others (\textit{Comfort with others}); (3) This notification mechanism disturbs my daily tasks (\textit{Disturbing}); (4) With this notification mechanism, I have the feeling that I am not missing an urgent notification anymore (\textit{Not missing notifications}); (5) This notification mechanism provides me the information that I want (\textit{Providing wanted information}). Furthermore, to evaluate notifications within the immersive environment, we adapted questions from Ghosh et al.'s work~\cite{ghosh2018notifivr} to assess noticeability, understandability, and perceived hindrance. The original questionnaire~\cite{ghosh2018notifivr} concerning \textit{Notification in VR} encompasses four scales: \textit{Intrusiveness}, \textit{Noticeability}, \textit{Understandability}, and \textit{Urgency}.  In this study, our focus was not on the specific placement of notifications but rather on implementing adapted designs that respond to different levels of urgency. Consequently, given that our system inherently categorised notifications by their urgency levels, we omitted the \textit{Perceived Urgency} scale from our adapted questionnaire to better align with our research objectives. We ended up with an adapted questionnaire about the \textit{Notification in VR} with 7-Likert scales as follows: (1) How much of a hindrance is the notification to your task (\textit{Intrusiveness}); (2) How easy or difficult is it to notice the notification you wanted (\textit{Noticeability}); (3) Once you notice the notification, how easy or difficult is it to understand what it stands for? (\textit{Understandability}). Additional questions captured overall system preferences across different scenarios and tasks, supplemented by open-ended items for qualitative feedback. For example, we asked the following questions: \textit{``Why do you prefer adaptive/default notification system?''} and \textit{``When using the adaptive notification system, would you be worried about missing important notifications?''}

\section{Results}

\begin{figure*}[!t]
    \centering
    \includegraphics[width=\linewidth]{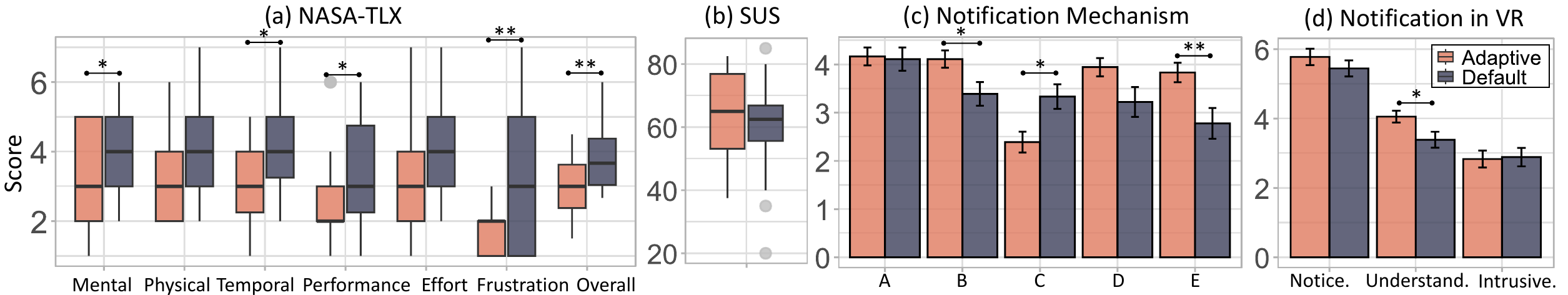}
    \caption{Average Scales of the (a) NASA-TLX~\cite{hart2006nasa}, (b) SUS~\cite{brooke1996sus}, (c) Notification Mechanism~\cite{weber2016design}, and (d) adapted Notification in VR~\cite{ghosh2018notifivr}. Note A for ``Not missing notifications'', B for ``Providing wanted information'', C for ``Disturbing'', D for ``Comfort'' and E for ``Comfort with others'', while Notice. is the abbreviation for ``Noticeability'', Understand. represents ``Understandabiliy'', and Intrusive. represents ``Intrusiveness''. ($p < 0.05$ (*), $p < 0.01$ (**)).}
    \label{fig:Evaluation/Data}
    \vspace{-5mm}
\end{figure*}

We conducted comprehensive quantitative analyses of both objective and subjective data to evaluate our system's performance. Paired-sample t-tests were employed for pairwise comparisons when data met normality assumptions. When the Shapiro-Wilk test indicated non-normal distribution $(p < 0.05$), we applied the Wilcoxon signed-rank test as a non-parametric alternative. Additionally, logistic regression was performed to examine potential correlations.

\subsection{Misclassification Impact on System Preference}

\begin{figure}[ht!]
    \centering
    \includegraphics[width=0.8\linewidth]{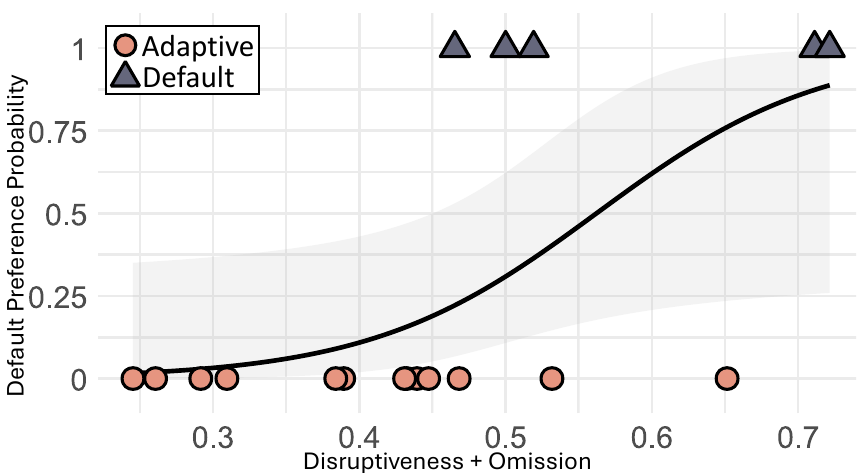}
    \caption{Logistic Regression Curve: Preference for Default vs. Disruptiveness + Omission}
    \label{fig:ClassificationAccuracy}
    \vspace{-5mm}
\end{figure}

To analyse the relationship between classification accuracy and user preferences, we defined two key error dimensions: \textit{Disruptiveness} and \textit{Omission}. These metrics capture the two primary ways notification systems can fail users. \textit{Disruptiveness} measures the percentage of \textit{not urgent} messages incorrectly classified as \textit{moderately urgent} or \textit{very urgent}, creating unnecessary interruptions. When treating \textit{non-urgent} notifications as negative cases, $Disruptiveness = \frac{FP}{FP + TN} = FPR$\footnote{FP: False Positive, TN: True Negative, and FPR: False Positive Rate}. \textit{Omission} represents the opposite error: the percentage of truly urgent messages (\textit{moderately urgent} and \textit{very urgent}) misclassified as \textit{not urgent}, causing users to potentially miss critical information. When treating \textit{moderately urgent} and \textit{very urgent} notifications as positive cases, $Omission = \frac{FN}{FN + TP} = FNR$\footnote{FN: False Negative, TP: True Positive, and FNR: False Negative Rate}.

\autoref{fig:ClassificationAccuracy} visualises these error rates for each participant. Each scatter point represents an individual participant, with position indicating their experienced \textit{Disruptiveness} and \textit{Omission} percentages during the study. The shape and colour of each point indicate the participant's stated notification preference.

To quantify how these error types influence user preferences, we performed logistic regression using both dimensions as predictors: $\log(\frac{p}{1-p}) = \beta_0 + \beta_{1}\times(Disruptiveness + Omission) = \beta_0 + \beta_1 \times (FNR + FPR)$. The logistic regression model yielded the coefficients: The intercept was 7.27 ($z = 2.201, p = 0.028$) and the coefficient for the combined predictor $\textit{Disruptiveness} + \textit{Omission}$ was -12.93 ($z = -1.996$, $p = 0.046$). The log-likelihood ratio test showed that the model with the combined predictor significantly improved fit over the intercept-only model, $\chi^{2}(1) = 6.92,\; p = 0.0085$ and McFadden's pseudo $R^{2} = 0.325$. The significant relationship between classification accuracy ($FPR+FNR$) and user preference establishes a critical design threshold for adaptive notification systems in MR. Users' tolerance for classification errors indicates that spatial distribution approaches provide benefits that offset AI imperfections. This finding may have crucial deployment implications: adaptive systems need not achieve perfect accuracy to deliver value, but developers should focus on minimising combined error rates. The predictive model offers a framework for evaluating and tuning classification algorithms before deployment.

\subsection{Questionnaire Results}
In this section, we reported the subjective data, including assessments using the \textit{Raw NASA-TLX}, \textit{SUS}, and a series of questions related to \textit{Notification in VR}~\cite{ghosh2018notifivr} and \textit{Notification Mechanism}~\cite{weber2016design}. Since the analysis of SUS scores revealed no statistically significant differences between the two systems, we did not discuss this further in this section.

\subsubsection{Raw NASA-TLX} 

\autoref{fig:Evaluation/Data} (a) shows the \textit{Raw NASA-TLX} ratings across all categories and the overall workload. Pairwise comparisons revealed that the \textit{default} system consistently demanded a higher workload across each category. In several dimensions, this difference reached statistical significance. Notably, the \textit{adaptive} notification system was associated with a significantly lower \textit{mental} workload ($Z = -2.092$, $p = 0.041$, $r = 0.493$), \textit{temporal} workload ($Z = -2.798$, $p = 0.008, r = 0.660$), \textit{frustration} ($Z = -2.9519$, $p = 0.004$, $r = 0.696$), \textit{performance} ($Z = -2.1342$, $p = 0.039$), and \textit{overall} workload ($Z = -2.841$, $p = 0.003$, $r = 0.670$). It is important to note that within the context of the \textit{Raw NASA-TLX}, a lower performance suggests superior system efficacy.

Significant reductions in mental workload ($p = 0.041$), temporal workload ($p = 0.008$), and frustration ($p = 0.004$) demonstrate that spatial notification distribution effectively addresses core MR usability challenges. The maintained or improved task performance indicates users can execute primary tasks more effectively while managing notifications. These findings validate that matching notification urgency to spatial attention demands reduces cognitive interference. For MR designers, this supports prioritising urgency-aware spatial layouts over centralised approaches, particularly in cognitively demanding applications such as productivity tools, training simulations, and data visualisation environments.



\subsubsection{Questionnaire about Notification Mechanism}
Notably, the Shapiro-Wilk normality test showed that measured differences were normally distributed in \textit{Comfort with others} ($W = 0.902$, $p = 0.062$), \textit{Disturbing} ($W = 0.903$, $p = 0.066$), and \textit{Providing wanted information} ($W = 0.909$, $p = 0.084$). Thus, we conducted paired-sample t-tests for pairwise comparisons. \autoref{fig:Evaluation/Data} (c) shows that the \textit{adaptive} system was rated significantly higher than the \textit{default} in \textit{Comfort with others} ($t(17) = 3.432$, $p = 0.003$, $\text{mean difference} = 1.06$) and \textit{Providing wanted information} ($t(17) = 2.718$ $p = 0.015$,$\text{ mean difference} = 0.722$), and significantly lower in \textit{Disturbing} ($t(17) = -2.878$, $p = 0.010$, $\text{ mean difference} = -0.944$).

The significantly higher ratings for \textit{Comfort with others} ($p = 0.003$) and \textit{Providing wanted information} ($p = 0.015$), combined with lower \textit{Disturbing scores} ($p = 0.010$), demonstrate that adaptive notifications better align with social and informational expectations in mixed reality contexts. The improved social comfort suggests that urgency-based spatial positioning appears less intrusive to bystanders, addressing a key barrier to MR adoption in shared spaces. The better information provision ratings indicate that users perceive the adaptive system as more intelligent and contextually aware, potentially increasing trust and long-term engagement with MR notification systems.

\subsubsection{Notification in VR Questionnaire}
Participants rated the \textit{Understandability} of the \textit{adaptive} notification system significantly higher than that of the \textit{default} ($Z = 2.396, \space p = 0.023,  \space r = 0.565$). Higher \textit{Understandability} scores ($p = 0.023$) indicate that users can better comprehend the meaning and content of individual notifications when they are positioned according to urgency rather than all placed centrally. This suggests that appropriate spatial placement reduces cognitive interference, allowing users to more effectively process and understand what each specific notification message represents, rather than struggling to interpret notification content while managing the distraction of inappropriate positioning.

\subsubsection{Further Questions} 

Overall, 13 participants preferred the adaptive notification system, while 5 participants favoured the default system. All participant responses were recorded in their native language and subsequently translated into English for analysis. Participants who preferred the adaptive system highlighted its ability to reduce disruptiveness while maintaining notification awareness. P12 noted: ``\textit{This new system perfectly addresses my dilemma of not wanting to use Do Not Disturb mode while also not wanting to see every single notification.}'' Several participants appreciated the control mechanisms, with P2 stating: ``\textit{I am very into the very urgent notification design as it provides a buffer for me to access the information which makes me feel much less disrupted.}'' Similarly, P4 emphasised the user agency: ``\textit{The moving notification for urgent messages gives me control that I can pull it forward when I'm ready.}'' The hand-anchored placement for moderately urgent notifications received positive feedback for its naturalness. P5 reported: ``\textit{Hand-anchored placement is natural. In reality, I also tend to check the notifications on my smartwatch.}'' However, some participants expressed concerns about potential notification oversight. Three participants (P3, P14, P16) reported feeling insecure about possibly missing notifications with the adaptive system. While P16 valued the reduced disruptiveness enough to still prefer the adaptive system overall, P3 and P14 found the risk of missing urgent notifications unacceptable, leading them to favour the default system despite acknowledging its higher disruptiveness. Additionally, P7 preferred the default system due to its consistent presentation format, noting that the dynamic configurations of adaptive notifications required higher cognitive processing than the predictable default positioning. Similarly, P15 reported feeling reduced control with the adaptive system, preferring to manually review all notifications and make individual response decisions rather than relying on automated urgency classification.

\section{Discussion and Limitation} 
\label{Discussion}

\subsection{Key Findings and Implications}

Our empirical findings demonstrate the effectiveness of adaptive notification UIs in MR environments. Analysis of the correlation between classification accuracy and user preferences (See~\autoref{fig:ClassificationAccuracy}) revealed a statistically significant pattern: participants favoured the adaptive system when the combined misclassification rate (Disruptiveness + Omission) remained below the threshold. This confirms hypothesis \textbf{[H1]} that users prefer the adaptive notification system over the default system even with classification errors present. The results further support \textbf{[H2]}, as our model demonstrates that user preference decreases as the combined error rate increases, indicating that participants were more likely to prefer the adaptive system when classification accuracy was higher. However, individual differences were evident, as users may exhibited strong preferences independent of misclassification rate. For example, P15 reported preferring to manually review all incoming notifications to ensure no very urgent messages were missed, prioritising completeness over reduced interruption. This pattern suggests that while the adaptive system benefits most users within acceptable error bounds, individual notification management strategies and risk tolerance vary considerably. Future systems may need to accommodate both adaptive and user-controlled notification modes to address diverse user preferences and contexts where missing critical information carries unacceptable consequences.


The quantitative results from NASA-TLX measurements show significantly lower mental workload ($p = 0.041$), temporal workload ($p = 0.008$), and frustration level ($p = 0.004$) with the adaptive notification system compared to the default system. These findings strongly support hypothesis \textbf{[H3]} that participants would experience measurably lower cognitive workload with the adaptive position-based MR notification system. This reduced cognitive load can be attributed to the spatial distribution of notifications based on urgency levels, which aligns with users' natural attentional priorities, a key benefit identified for our adaptive notification system design.

Notification awareness remained comparable between systems, supporting hypothesis \textbf{[H4]}, with the adaptive approach demonstrating superior information processing capabilities. Participants rated the adaptive system significantly higher for \textit{Providing wanted information} ($t(17) = 2.718$, $p = 0.015$) and \textit{Understandability} ($Z = 2.396$, $p = 0.023$). These findings demonstrate that position-based urgency mapping effectively reduces cognitive interference while enhancing users' ability to perceive and comprehend notification content, even when urgency classifications contain errors. This may be attributed to the adaptive system's ability to reduce cognitive overload and optimise information processing. On the contrary, the default system creates cognitive overload by displaying all notifications centrally, regardless of urgency, forcing users to process every message with equal attention. This constant interruption impairs users' ability to process information effectively~\cite{horvitz2001notification, ohly2023effects} and degrades their capacity to distinguish critical information from routine updates, resulting in lower \textit{Understandability} scores and increased risk of missing important notifications. Conversely, the adaptive system's spatial urgency mapping enables contextual information processing. Peripheral placement of non-urgent items reduces immediate cognitive burden, hand-anchored moderately urgent notifications provide awareness without disruption~\cite{rzayev2019notification, ghosh2018notifivr}, and gradual transition of very urgent messages allows users to mentally prepare for important content. This structured approach prevents information overload, enables better attention allocation, and allows users to engage with notifications when cognitively ready, ultimately enhancing both comprehension and perceived information value~\cite{arnold2023dealing}.

Beyond our hypotheses, we observed that the default notification system altered participants' response behaviours: users reported responding to messages they would normally consider non-urgent when presented in the centralised format. This suggests that constant central placement creates artificial urgency, compelling users to engage with low-priority content. We term this phenomenon as ``pseudo-urgency''. This misallocation of attention may lead to frequent task interruptions and reduced efficiency in primary activities, which can compound over time and prove particularly problematic in work environments requiring sustained focus~\cite{ohly2023effects, stothart2015attentional}.

\subsection{Practical Implications for MR/VR Notification Design}

\paragraph{Spatial Design Principles for MR Notification Systems}
Our findings challenge the centralised notification placement used in commercial MR headsets~\cite{10.1145/3467963} and provide actionable design guidelines for more effective approaches. Unlike the previous work which focused on the static notification placement~\cite{rzayev2019notification, hsieh2020bridging} and adaptive notification placement based on gaze contexts~\cite{kawakubo2025dynamic, GZ2024Ilo}, our approach incorporates urgency-based classification to determine where and how notifications should appear, providing a more comprehensive adaptive framework for MR environments. The urgency-based spatial positioning significantly outperformed centralised approaches, validating and extending Rzayev et al.'s static placement work~\cite{rzayev2019notification} by demonstrating the benefits of dynamic adaptation. Based on these results, MR designers should implement spatial hierarchies where non-urgent notifications remain peripheral, moderately urgent content anchors to hands, and critical information transitions progressively to the central view.

\paragraph{Leveraging Embodied Cognition for Notification Design}

The hand-anchored placement for moderately urgent notifications builds on embodied cognition principles demonstrated by Ens et al.~\cite{ens2014cockpit} and Grubert et al.~\cite{grubert2015multifi}. This approach reduces disruptive gaze shifts while maintaining awareness, directly addressing attention management challenges identified by Ghosh et al.~\cite{ghosh2018notifivr} and supporting perceptual load theory~\cite{lavie1995perceptual}. The significant user preference for hand-anchored notifications over head-up displays found by Rzayev et al. further validates leveraging the body as an integral notification component in immersive environments~\cite{rzayev2019notification}. However, we acknowledge that hand-anchored notifications have inherent limitations. In contexts involving continuous hand movement, such as rhythm games, notifications may temporarily overlap with task elements and affect users' performance~\cite{hsieh2020bridging}. Conversely, during hands-free activities such as passive observation or locomotion, reduced hand monitoring may diminish notification visibility. However, these limitations align with the characteristics of moderately urgent notifications, which are deferrable by nature and do not require immediate acknowledgement. The design deliberately prioritises reduced disruption over guaranteed visibility, making occasional missed notifications an acceptable trade-off.

\paragraph{Progressive Disclosure and User Agency}

The progressive disclosure mechanism for very urgent notifications implements Calm Technology principles~\cite{weiser1996designing} while addressing immersion-breaking concerns raised by Hsieh et al.~\cite{hsieh2020bridging}. This two-stage attention mechanism balances immediate awareness with controlled interruption. Together, these design elements demonstrate that effective MR notification systems must respect user attention as a scarce cognitive resource while providing appropriate control mechanisms.

\subsection{Limitations}
Several limitations should be considered when interpreting our findings. While prioritising ecological validity improves real-world relevance, it introduces threats to internal validity by increasing uncontrolled variance. The use of multiple environments and tasks introduces confounding factors such as differing cognitive demands, engagement levels, and environmental salience, which may independently affect workload and notification perception, thereby increasing measurement noise. Additionally, our notification delivery rate exceeded typical real-world patterns. While this accelerated pace balances reasonable experiment duration, thus minimising participant fatigue, with the real world, it may have amplified cognitive load effects and influenced user preferences in ways that diverge from authentic usage scenarios. Furthermore, our classification methodology represents a constraint, as we used pre-labelled notifications rather than real-time AI classification systems. While this approach protected participant privacy and ensured controlled experimental conditions, it may not fully capture the dynamic challenges of automated urgency detection deployed in practice. Real AI systems typically require extensive user data and interaction history~\cite{mehrotra_prefminer_2016, pielot2017beyond, zheng2025persono}, introducing additional classification errors that our controlled study design could not replicate. Additionally, our participant sample consisted predominantly of VR-inexperienced users, which may limit the generalisability of our findings to expert VR users who might have different expectations and interaction patterns with notification systems in immersive environments. Furthermore, our participant demographics were unevenly distributed, with 14 females and only 4 males, which may affect the generalisability of our findings given potential gender differences in notification preferences and interruption tolerance.

Technical constraints also affected our implementation. The Quest 3's lack of eye-tracking capabilities required reliance on HMD-mounted cameras for gaze approximation, leading to occasional recognition failures when users read notifications without directly facing the cameras. This necessitated gesture controls as backup interaction methods and may have influenced user preferences and workload measurements. Additionally, our VR-based Mixed Reality Simulation approach~\cite{lee2010role, li2019gaze, lu2022exploring}, while providing experimental control, cannot fully replicate the environmental complexity and social dynamics of authentic MR usage scenarios in diverse physical environments.

Last, ecological validity concerns arise from our controlled laboratory environment and standardised notification content. Real-world notification streams vary significantly in content, sender relationships, and contextual relevance, which could affect urgency perception and classification accuracy. The artificial nature of our tasks, while representing common VR use cases, may not capture the complexity and personal investment of authentic MR activities, where notification management becomes critical for user adoption and satisfaction.

\subsection{Toward Real-World Deployment and Future Work}

Our logistic regression analysis demonstrates that AI-driven adaptive notification systems need not achieve perfect accuracy (100\%), an impossibility established by prior work~\cite{li2023alert, zheng2025persono}, to earn user preference. This finding contrasts with previous research showing that imperfect machine learning-based AR cues significantly degrade performance in visual search tasks, where cues provide task-critical guidance and classification errors directly misdirect attention and action~\cite{Raikware2025Beyond}. However, notification systems operate fundamentally differently: urgency is inherently subjective and context-dependent, making perfect inference unattainable and prompting users to accept some degree of imperfection. Our results indicate that notifications serve as attention-management aids rather than directive task cues, enabling the benefits of adaptive spatial presentation to outweigh the costs of moderate misclassification.

Since our study was conducted under controlled laboratory conditions, and real-world deployment faces substantial challenges. In our study, notifications appeared sequentially at randomised intervals. In practice, users may receive multiple simultaneous messages or notifications with varying urgency levels in one conversation. Naively repositioning each message individually risks spatial instability and increased cognitive load. Possible strategies include fixing a conversation to a stable spatial anchor while modulating salience, aggregating multiple notifications of similar urgency (e.g., not urgent notifications from the same application are clustered in the same icon), or smoothing urgency transitions over time to avoid abrupt spatial shifts. These approaches suggest that urgency should be treated as a dynamic signal with temporal continuity rather than a discrete per-message attribute, warranting further investigation in longitudinal deployments. Additionally, all primary tasks in our study were hand-based. In real-world scenarios, some tasks may not require hand interaction; therefore, on-hand notification design requires further exploration in practical deployment contexts.

Moreover, our study revealed substantial individual variation in notification management preferences. Some users preferred comprehensive control, manually reviewing all notifications, while others trusted automated system classifications. Therefore, future deployment of adaptive notification systems should adhere to Human-Centered Artificial Intelligence principles~\cite{shneiderman2020human}, balancing automation with user agency to accommodate diverse preferences. Several design strategies could achieve this balance. First, systems could provide user-configurable interfaces similar to DataHalo~\cite{han2023datahalo}, enabling users to define custom notification rules and keywords (e.g., classifying all messages from their supervisor as very urgent). Notifications not matching predefined rules would default to automated urgency assessment. Second, urgency control could be partially delegated to senders, as demonstrated by Cho et al.~\cite{cho2020share}, allowing message originators to signal time-sensitivity explicitly. These hybrid approaches would enable personalisation while maintaining the cognitive benefits of adaptive spatial positioning for uncategorized notifications.

Finally, another critical next step involves deploying our adaptive notification system on actual MR headsets when hardware capabilities reach sufficient maturity.  Current limitations in eye-tracking precision, gesture recognition, and real-time processing power on consumer MR devices constrain immediate implementation. However, as these technical barriers are addressed, future work should prioritise transitioning from controlled laboratory environments to authentic deployment scenarios. More importantly, our study did not examine the impact of misclassifying urgent notifications as non-urgent, which can only be revealed through real-world deployment. Therefore, future work should explore whether users maintain the preference patterns identified in our study after experiencing the actual consequences of misclassification. Furthermore, longitudinal field studies spanning multiple months will be essential to validate our findings in real-world contexts. Such studies should examine how user preferences and system effectiveness evolve over extended usage periods, addressing questions our short-term evaluation could not capture: Do users maintain their preference for adaptive systems after weeks of daily use? How do habituation effects influence the perceived benefits of urgency-based spatial positioning? What adaptation strategies do users develop when classification errors occur repeatedly in their personal notification streams?


\section{CONCLUSION}
This research demonstrates that adaptive notification systems in mixed reality significantly improve user experience by reducing cognitive workload while maintaining notification awareness. Our evaluation with 18 participants reveals that urgency-based spatial positioning, featuring peripheral non-urgent notifications, hand-anchored moderately urgent messages, and progressively transitioning very urgent content. It outperforms centralised approaches across mental workload, temporal demand, and frustration metrics. Users prefer adaptive systems even with classification errors, provided combined misclassification rates remain below our regression-determined threshold. These findings challenge current commercial MR designs and establish that developers should prioritise balancing disruptiveness and omission errors over pursuing perfect classification accuracy, while implementing spatial hierarchies that treat user attention as a finite cognitive resource.
\newpage 

\acknowledgments{This research is supported by the Hong Kong Polytechnic University (PolyU)'s Start-up Fund for New Recruits Under Grant (Project ID: P0046056), PolyU Department of Industrial and Systems Engineering Under Grant (Project ID: P0056354), and the Research Grant Council's General Research Fund Under Grant (Project ID: P0056902). Jingyao Zheng and Xian Wang were supported by a grant from the PolyU Research Committee under student account codes RMCU and RMHD, respectively. This work has been partly supported by the Research Center Trustworthy Data Science and Security (\href{https://rc-trust.ai}{https://rc-trust.ai}), one of the Research Alliance centers within the UA Ruhr (\href{https://uaruhr.de}{https://uaruhr.de}).}

\bibliographystyle{abbrv-doi-hyperref}

\bibliography{bibliography.bib}

\end{document}